\documentclass[aps,pra,superscriptaddress,reprint,floatfix,longbibliography,showpacs,amsfonts,amsmath,amssymb]{revtex4-1}
\usepackage{graphicx}
\usepackage{hyperref}
\usepackage{xr-hyper}
\usepackage{upgreek}
\usepackage{epstopdf}
\usepackage[note-name=, use-sort-key = false]{notes2bib}
\usepackage{color}
\usepackage{comment}

\begin{document}
\title{Chemical pressure effect on the optical conductivity of the
nodal-line semimetals ZrSi$Y$ ($Y$=S, Se, Te) and ZrGe$Y$ ($Y$=S, Te)}

\author{J. Ebad-Allah}
\affiliation{Experimentalphysik II, Augsburg University, 86159 Augsburg, Germany}
\affiliation{Department of Physics, Tanta University, 31527 Tanta, Egypt}
\author{J. Fern\'andez Afonso}
\affiliation{Institute of Solid State Physics, TU Wien, 1020 Vienna, Austria}
\author{M. Krottenm\"uller}
\affiliation{Experimentalphysik II, Augsburg University, 86159 Augsburg, Germany}
\author{J. Hu}
\affiliation{Department of Physics, University of Arkansas, Fayetteville, AR 72701, USA}
\author{Y. L. Zhu}
\affiliation{Department of Physics, Pennsylvania State University, University Park, PA 16803, USA}
\affiliation{Department of Physics and Engineering Physics, Tulane University, New Orleans, LA 70118, USA}
\author{Z. Mao}
\affiliation{Department of Physics, Pennsylvania State University, University Park, PA 16803, USA}
\affiliation{Department of Physics and Engineering Physics, Tulane University, New Orleans, LA 70118, USA}
\author{J. Kune\v{s}}
\email{kunes@ifp.tuwien.ac.at}
\affiliation{Institute of Solid State Physics, TU Wien, 1020 Vienna, Austria}
\affiliation{Institute of Physics, The Czech Academy of Sciences, 18221 Praha, Czech Republic}
\author{C. A. Kuntscher}
\email{christine.kuntscher@physik.uni-augsburg.de}
\affiliation{Experimentalphysik II, Augsburg University, 86159 Augsburg, Germany}

\begin{abstract}
ZrSiS is a nodal-line semimetal, whose electronic band structure contains a diamond-shaped line of
Dirac nodes. We carried out
a comparative study on the optical conductivity of ZrSiS and the related compounds ZrSiSe, ZrSiTe, ZrGeS, and ZrGeTe by reflectivity measurements over a broad frequency range combined with density functional
theory calculations. The optical conductivity exhibits a distinct U-shape, ending at a sharp peak at around 10000~cm$^{-1}$
for all studied compounds, except for ZrSiTe. The U-shape of the optical conductivity is due
to transitions between the linearly dispersing bands crossing each other along the nodal line.
The sharp high-energy peak is related to transitions between almost parallel bands, and its energy position depends on the interlayer bonding correlated with the $c$/$a$ ratio, which can be tuned by either
chemical or external pressure. For ZrSiTe, another pair of crossing bands appears in the vicinity of the
Fermi level, corrugating the nodal-line electronic structure and leading
to the observed difference in optical conductivity.
The findings suggest that the Dirac physics in Zr$XY$ compounds with $X$=Si, Ge and $Y$=S, Se, Te is
closely connected to the interlayer bonding.
\end{abstract}

\maketitle

\section{Introduction}

The quest for novel topological materials with exceptional properties has led in recent years to an enormous research activity
on two-dimensional (2D) and three-dimensional (3D) Dirac semimetals hosting massless Dirac fermions. In Dirac semimetals
linearly dispersing bands cross at isolated points
at the Fermi energy $E_F$,
resulting in a dispersion locally resembling that of massless fermions.
The most popular example for a 2D Dirac semimetal is graphene exhibiting highly interesting properties, such as outstanding mechanical
stability \cite{Lee.2008}, ultrahigh electron mobility \cite{Bolotin.2008}, and superior thermal conductivity \cite{Balandin.2008}.
In contrast to Dirac semimetals with discrete Dirac points or nodes, in nodal-line semimetals the linearly dispersing bands cross along a line in the Brillouin zone. Corrugation of the
nodal lines in the energy direction then gives rise to rod-shaped Fermi surfaces, which are sensitive to small changes in external
parameters.

ZrSiS is a nodal-line semimetal with a diamond-rod-shaped Fermi surface
resulting from an almost ideal nodal-line band structure \bibnote{Please note that there are additional surface states in ZrSiS and related compounds
according to angle-resolved photoemission studies combined with band structure calculations \cite{Neupane.2016,Schoop.2016,Topp.2017}. However, these additional
surface states are not relevant for our optical study, which probes the bulk electronic properties.}.
It was suggested that its band structure contains additional
Dirac-like band crossings located several hundred meV above and below $E_F$, which are protected by non-symmorphic
symmetry \cite{Neupane.2016,Schoop.2016,Topp.2016}. ZrSiS is exceptional among the Dirac materials since the linearly dispersing
bands at $E_F$ extend over a rather large energy range of up to 2~eV, although spin-orbit coupling introduces a small gap ($\sim$0.02~eV)
to the Dirac nodes. It shows exceptional properties like
unusual magnetoresistance characteristics \cite{Lv.2016,Sankar.2017}, high mobility of charge carriers \cite{Sankar.2017,Hu.2017}
and correlation effects \cite{Pezzini.2017}, which renders ZrSiS a highly interesting material.

ZrSiS belongs to the family of compounds Zr$XY$, where $X$ can be a carbon group element ($X$=Si, Ge, Sn) and $Y$ is a chalcogen
element ($Y$=S, Se, Te) \cite{Wang.1995}. In this compound family the 2D--to--3D structural dimensionality evolution induced by
isoelectronic substitution, which is generally called chemical pressure, was suggested to be realized and to be monitored $via$
the ratio of lattice parameters $c$/$a$ \cite{Wang.1995,Klemenz.2018}. This renders Zr$XY$ a model system to probe the effect of structural dimensionality change in nodal-line semimetals.

Earlier optical studies on ZrSiS found a rather flat optical conductivity in the energy range 30 -- 300~meV, which was claimed
to be due to 2D Dirac bands near E$_F$ \cite{Schilling.2017}. This interpretation was, however,
questioned recently by theoretical calculations \cite{Habe.2018}. The high-energy optical conductivity
exhibits a distinct U-shape ending at a sharp kink/peak around 10000~cm$^{-1}$,
which was not addressed previously \cite{Schilling.2017}.

We present a combined experiment+theory study of several members of the Zr$XY$ family, which addresses these features
and relate them to the nodal-line electronic structure of the materials. In particular, we find that
ZrSiS, ZrSiSe, ZrGeS, and ZrGeTe have are very close to ideal nodal-line band structure
with varying degree of corrugation of the nodal line, whereas in ZrSiTe another
pair of crossing bands appears in the vicinity of the Fermi level, destroying the nodal-line structure and pushing it away from $E_F$ in parts of the Brillouin zone.

\begin{figure}[tb]
\includegraphics[width=0.45\textwidth]{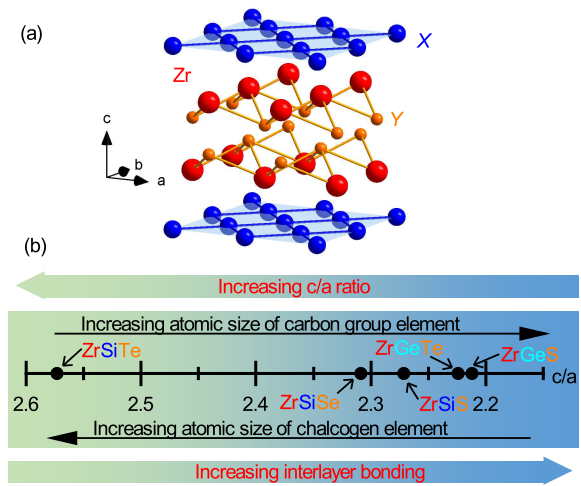}
\caption{(a) Crystal structure of the Zr$XY$ compounds with square nets parallel to the $ab$ plane.
(b) Scheme illustrating the chemical pressure effect in ZrSiS, ZrSiSe, ZrSiTe, ZrGeS, and ZrGeTe.}\label{figure1}
\end{figure}

The compounds Zr$XY$ crystallize in a PbFCl-type structure in the tetragonal P4/nmm space group with non-symmorphic
symmetry [see Fig.\ \ref{figure1}(a)] \cite{Wang.1995}. The layered crystal structure contains square nets of $X$
atoms parallel to the $ab$ plane, where each $X$ square net is sandwiched between two Zr square nets. Each $X$ atom has bonding
to four Zr atoms in tetrahedral coordination. The chalcogen atoms $Y$ also form square nets in the $ab$ plane. Hence, the crystal
structure of Zr$XY$ consists of slabs with five square nets with a stacking sequence [$Y$-Zr-$X$-Zr-$Y$], terminated by $Y$ square
nets on both sides \cite{Sankar.2017}. The bonding between two adjacent slabs is of van der Waals type and hence weak. Therefore,
these crystals generally tend to cleave along the plane between two chalcogen layers.

It was proposed that the structural dimensionality of the Zr$XY$ materials can be tuned by chemical pressure, i.e., isoelectronic substitution, and monitored $via$ the ratio of lattice
parameters $c$/$a$, where $c$ is the distance between two adjacent Si square nets and $a$ is the in-plane lattice parameter
of the tetragonal crystal structure \cite{Wang.1995,Klemenz.2018}. The $c$/$a$ ratio may serve as a measure for the interlayer bonding strength in the system.
As illustrated in Fig.\ \ref{figure1}(b), the chemical pressure effect in the studied Zr$XY$ compounds can be realized either by the
isoelectronic substitution of the carbon group element $X$ or the chalcogen element $Y$.
With increasing the atomic size of the chalcogen element $Y$ the slab thickness along the $c$ axis increases and hence the $c/a$ ratio increases, causing a decrease in the interlayer bonding.
Accordingly, the chemical pressure in ZrSiTe is reduced as compared to ZrSiS, and therefore ZrSiTe was described as a layered material
without significant interlayer bonding \cite{Wang.1995}.
In contrast, with increasing atomic size of the carbon
group element $X$ the $c/a$ ratio decreases, and thus tunes the interlayer bonding in the opposite way.
For example, in ZrSiS a smaller interlayer bonding is expected as compared to ZrGeS [see Fig.\ \ref{figure1}(b)].
In the case of the $X$ substitution, the chemical pressure effect is, however, less direct as described in Ref.\ \onlinecite{Wang.1995}: The increasing atomic size of $X$ leads to longer $X$-$X$ bonds within the $X$ square nets, which affects
the in-plane $Y$-$Y$ spacing as well. The cell parameter $c$ is decreased, whereas the unit cell expands along the $a$ direction. The net effect is an enhancement of the interlayer bonding in adjacent [$Y$-Zr-$X$-Zr-$Y$] slabs.

Among all studied materials, ZrSiTe has the highest $c$/$a$ ratio, with a value close to 2.6, and therefore supposedly has a smaller interlayer bonding.
This is confirmed by de Haas-van Alphen quantum oscillation measurements, which show that the Fermi surface
has a 2D character in ZrSiTe, in contrast to ZrSiS and ZrSiSe ($c$/$a$ ratio close to 2.3) with a 3D-like Fermi surface \cite{Hu.2016}. This also explains why ZrSiTe single crystals can be easily exfoliated mechanically.

\section{SAMPLE PREPARATION, EXPERIMENTAL AND COMPUTATIONAL DETAILS}

Single crystals ZrSi$Y$ ($Y$=S, Se, Te) and ZrGe$Y$ ($Y$=S, Te) were grown by a chemical vapor transport method \cite{Hu.2017a,Hu.2017}.
The samples were characterized by x-ray diffraction and energy-dispersive x-ray spectroscopy, in order to ensure phase-purity and crystal quality.

The reflectivity  measurements were carried out in the frequency range ($\approx$ 100 -- 24000~cm$^{-1}$) with an infrared microscope (Bruker Hyperion), equipped with a 15$\times$ Cassegrain objective, coupled to a Bruker Vertex v80 Fourier transform infrared spectrometer. Reflectivity measurements on ZrSiSe, ZrSiTe, and ZrGeTe were performed on freshly cleaved (001) surfaces. Since ZrSiS and ZrGeS crystals cannot be easily cleaved, the as-grown, shiny (001) surfaces were carefully cleaned with isopropanol before the reflectivity measurements. The reproducibility of the results was checked on several crystals for each compound. To obtain the absolute reflectivity spectra, the intensity reflected from an Al mirror served as reference. The reflectivity measurements were carried out at room temperature, since the dependence of the optical response on temperature is relatively weak, especially regarding the interband transitions, as was shown for ZrSiS \cite{Schilling.2017}.

The frequency-dependent dielectric function $\epsilon(\omega)$ and optical conductivity $\sigma(\omega)$ of the materials were obtained via Kramers-Kronig analysis of the reflectivity spectra. To this end, the reflectivity data were extrapolated to low frequencies based on the Drude-Lorentz fitting, whereas x-ray atomic scattering functions \cite{Tanner.2015} were used for calculating the higher-frequency extrapolations. A power law 1/$\omega^{n}$ with $n$ up to 3 was used for interpolating the reflectivity spectra between the measured and calculated data. The contributions to the optical conductivity spectra $\sigma_1$ were obtained by simultaneous Drude-Lorentz fitting of the reflectivity and optical conductivity.

Pressure-dependent reflectance measurements up to 3.5 GPa at room temperature were carried out with an infrared microscope (Bruker Hyperion) coupled to a Bruker Vertex v80 Fourier transform infrared spectrometer in the frequency range 600-23000~cm$^{-1}$.
%The measurements were carried out on the same crystal piece.
All the pressure-dependent reflectivity spectra refer to the absolute reflectivity at the sample-diamond interface, denoted as $R_\text{sd}$. The reflectivity spectrum $R_\text{sd}(\omega)$ was calculated according to $R_\text{sd}(\omega)$= $R_\text{dia}\times I_\text{s}(\omega)/I_\text{ref}(\omega)$,
where $I_\text{s}$ is the intensity of the radiation reflected from the sample-diamond interface, $I_\text{ref}$ is the intensity reflected from the inner diamond-air interface of the empty diamond anvil cell, and $R_\text{dia}$ =0.167, which is assumed to be pressure independent \cite{Eremets.1992}.

The pressure-dependent reflectivity spectra are affected in the frequency range between 1800 and 2670~cm$^{-1}$ by multi-phonon absorptions in the diamond anvils, which are not completely corrected by the normalization procedure. Therefore, this part of the spectrum was interpolated based on the Drude-Lorentz fitting. The measured reflectivity data were extrapolated to low frequencies based on the Drude-Lorentz fit for further analysis. For the high-frequency extrapolation the simulated free-standing reflectivity spectrum (see above) was used, taking into account the sample-diamond interface. The real part of the optical conductivity $\sigma_{1}$ was obtained via Kramers Kronig transformation of the reflectivity spectrum \cite{Pashkin.2006}.

Density functional theory (DFT) calculations have been performed using the Wien2k~\cite{wien2k} code.
The calculations were carried out with the Generalized Gradient Approximation as exchange correlation functional, 1000 k-points in the self-consistent calculation
and 200000 k-point to evaluate the optical conductivity.
%Inclusion of spin-orbit coupling in the DFT Hamiltonian modifies/opens the gap close to the Fermi level, but it has no discernible effect on the optical spectra
%above 3000~cm$^{-1}$, which is the subject of the present paper.

In order to gain insight into the origin of the studied band structures we have constructed a tight-binding model in the basis of Zr-$d$, Si-$s,p$ and Y-$p$ Wannier functions. To this end we have used wannier90~\cite{wannier90} and wien2wannier~\cite{wien2wannier} packages.

\begin{figure}[t]
\includegraphics[width=0.5\textwidth]{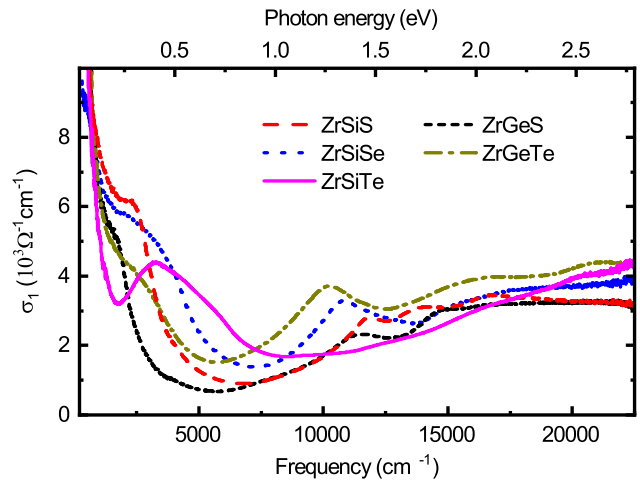}
\caption{Experimental optical conductivity spectra for all measured Zr$X$$Y$ compounds as obtained from ambient-condition reflectivity spectra.}\label{figure2}
\end{figure}

\section{RESULTS}

\begin{figure*}[t]
\includegraphics[width=0.95\textwidth]{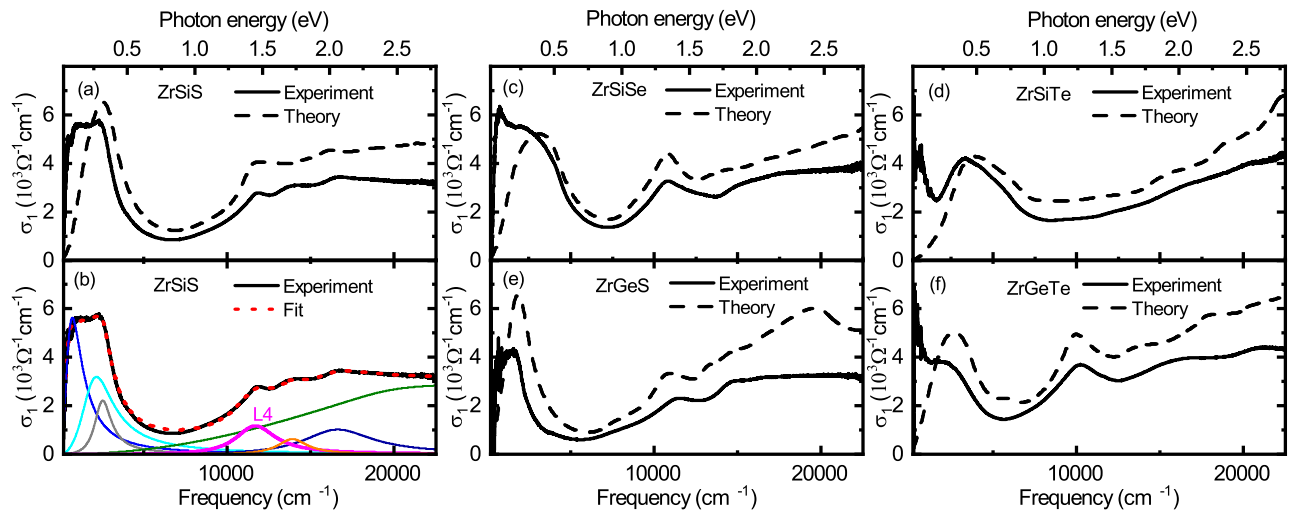}
\caption{(a), (c)-(f): Comparison of the experimental and theoretical optical conductivity without
intraband contributions for ZrSiS, ZrSiSe, ZrSiTe, ZrGeS, and ZrGeTe, respectively. (b) Contributions to the experimental
optical conductivity of ZrSiS as obtained from the fitting are shown, in particular the sharp peak labelled L4, which is due to transitions between almost parallel bands.}\label{figure3}
\end{figure*}

The real part of the optical conductivity $\sigma_1(\omega)$ of ZrSiS [see Fig.\ \ref{figure2} and suppl. Fig.\ S2(a)]
contains two Drude contributions, which is consistent with recent reports of the coexistence of electron-type and hole-type
charge carriers, where electron-hole compensation was suggested to cause and extremely large magnetoresistance \cite{Lv.2016,Singha.2017}.
Besides the Drude contributions, the low-energy optical conductivity shows a plateau-like behavior in the range
$\sim$240 -- 2400~cm$^{-1}$, which was claimed to be due to 2D Dirac bands near E$_F$ \cite{Schilling.2017}.
In analogy to graphene, in a 2D system with an ideal linear dispersion of Dirac bands one would expect a constant
optical conductivity \cite{Carbotte.2017,Mukherjee.2017,Ahn.2017}. The interpretation of the flat conductivity in
ZrSiS in terms of 2D Dirac bands was, however, questioned recently by theoretical calculations \cite{Habe.2018}.
The drop in the optical conductivity for higher frequencies and the subsequent rise result in a broad dip centered
at 6700~cm$^{-1}$, i.e., a U-shaped optical conductivity [see Figs.\ \ref{figure2} and \ref{figure3}(a)].
The high-energy optical conductivity of ZrSiS is rather flat, overlayed with three well-resolved absorption features,
which have not been considered previously. Among them, the sharp peak at 11650~cm$^{-1}$ [labelled L4 in
Fig.\ \ref{figure3}(b)] is the most pronounced.
A similar profile of the optical conductivity spectrum -- namely a distinct U-shape ending at a sharp peak -- is found
for the materials ZrSiSe, ZrGeS, and ZrGeTe (see Figs.\ \ref{figure2} and \ref{figure3}).
In particular, the high-energy L4 peak is present in all three compounds with slight variation in its energy position depending
on the carbon group element and the chalcogen element (chemical pressure).

In comparison to the other studied Zr$XY$ materials, the optical conductivity profile of ZrSiTe
is distinctly different [see Figs.\ \ref{figure2} and \ref{figure3}(d)]. Besides the Drude-like characteristics, the low-energy range is dominated by a pronounced absorption band centered
at $\approx$3300~cm$^{-1}$ with a shoulder on its high-frequency side [see suppl. Fig.\ S2(c)]. Above 9000~cm$^{-1}$
the optical conductivity monotonically increases with increasing frequency without any well-resolvable feature.
Most importantly, the sharp peak at around 10.000$^{-1}$ is no longer present in ZrSiTe.

\begin{figure*}[t]
\includegraphics[width=1\textwidth]{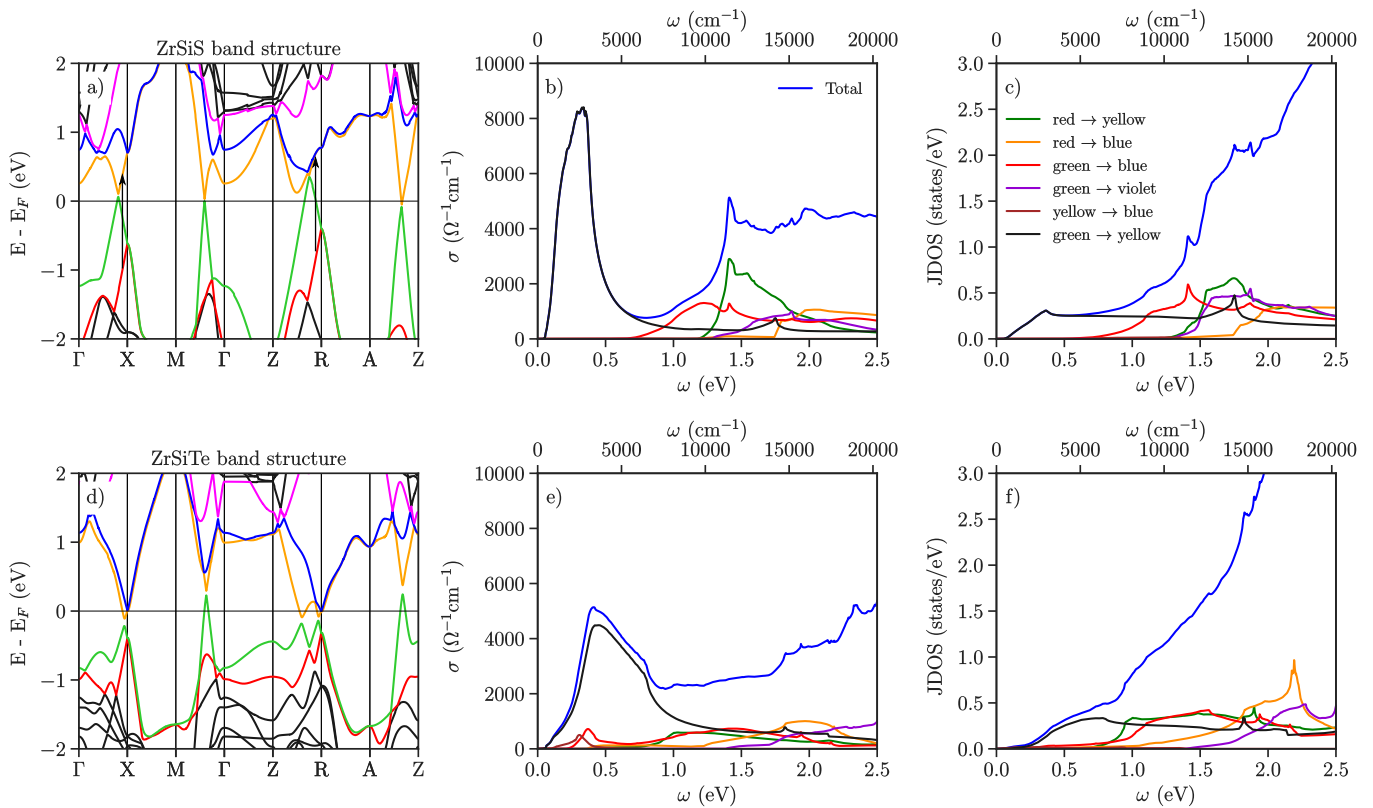}
\caption{Calculated band structures of ZrSiS (a) and ZrSiTe (d) with SOC, panels (b, e) and (c, f) show contributions of different band combinations to the optical conductivity $\sigma$ and the joint density of states JDOS, respectively. Black arrows in panel (a) highlight the
transitions between almost parallel red and yellow bands
in the vicinity of the $X$ and $R$ points, which give rise to the L4 peak in the optical conductivity.} \label{figure4}
\end{figure*}

\begin{figure*}[ptb]
\includegraphics[width=0.9\textwidth]{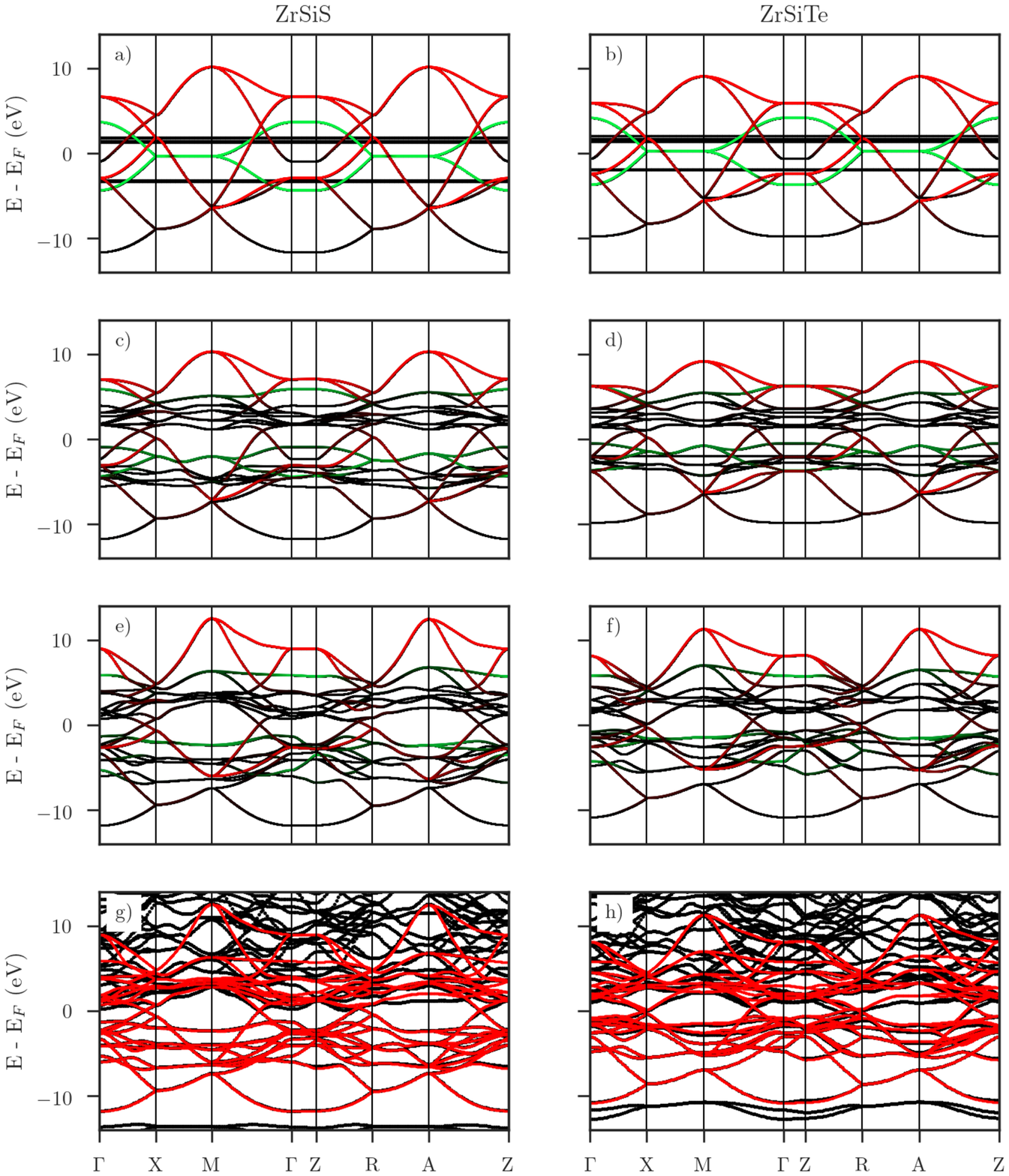}
\caption{Wannier band structures of ZrSiS (left) and ZrSiTe (right). Red and green colors in panels (a)-(f) reflect the Si $pxpy$ and Si $pz$ weights, respectively.
The panels (a)-(f) show the band structures obtained for hopping constrained to: nearest neighbors in the Si plane  (a)-(b), nearest neighbors Si-Si, Si-Zr and Zr-S/Te (c)-(d) and no constraints (e)-(f).
Panels (g,h) show the same Wannier bandstructure (red) as panels (e,f) compared to the full DFT band structures (black).}
\label{figure5}
\end{figure*}

\subsection{Theoretical optical spectra}
In order to understand the measured optical conductivities we have performed density functional
calculations with Wien2k~\cite{wien2k} code for obtaining the electronic band structure and optical spectra.
Here, we focus on the high-energy features in the optical spectra
$>3000$~cm$^{-1}$ and thus we did not consider the Drude intra-band contributions. The spin-orbit coupling plays no role in the optical conductivity
in this frequency range. Starting from the scalar-relativistic (no spin-orbit coupling) band structure, the spin-orbit coupling opens or enhances a small gap between the crossing bands forming the nodal line in the vicinity of the Fermi level. It thus affects the $dc$ transport and low-energy optical conductivity, however, being restricted to a low-dimensional subspace (vicinity of the nodal-line) of the Brillouin zone, the spin-orbit splitting was found to have no discernible effect on the optical conductivity in the frequency range $>3000$~cm$^{-1}$.

The comparison in Fig.~\ref{figure3} reveals a good quantitative agreement between experiment
and theory for all studied compounds.
With ZrSiTe being a clear outlier, all the remaining compounds exhibit a U-shaped optical conductivity between 3000-10000~cm$^{-1}$, bounded on the low-energy side by a flat region and on the high-energy side by a sharp peak
(see Fig.~\ref{figure3}).
In order to understand the origin of these features, we show the calculated band structure together with a decomposition of
the optical conductivity and joint density of states (JDOS) into contributions of different band combinations in Fig.~\ref{figure4}.
We point out that this decomposition is merely an analytic tool and does not have a deeper physical meaning.

First, we discuss the spectrum of ZrSiS. The dropping side of the U-shaped region of the optical conductivity spectrum ($<6000$~cm$^{-1}$) is dominated by
the transitions between the linear crossing bands close to $E_F$, marked green and yellow
in Fig.~\ref{figure4}(a). The rising side of the U-shaped region is due to other band combinations.
Analyzing the band and k-point contribution to the optical conductivity,
the sharp L4 kink marking the upper bound of the U-shaped region can be assigned to transitions between the almost parallel red and yellow bands
in the vicinity of the $X$ and $R$ points, as marked in Fig.~\ref{figure4}(a). This interpretation was first pointed out by Habe and Koshino~\cite{Habe.2018}. Therefore, the position of the kink in all the studied materials, except for ZrSiTe, follows the red/yellow splitting at the $X$ and $R$ points.

The comparison of the optical conductivity in Fig.~\ref{figure4}(b) with the JDOS in Fig.~\ref{figure4}(c) reveals an approximate $1/\omega$ relationship between the two quantities, which corresponds to constant momentum dipole matrix elements~\cite{Ahn.2017,Carbotte.2017} \bibnote{Please see suppl. Fig. S4
for further information regarding the importance of dipole matrix elements for the in-plane and out-of-plane optical conductivity of ZrSiS.}.
The JDOS originating from the lowest bands exhibits a broad constant region following the initial onset [black line in Fig.~\ref{figure4}(c)].
%consistent with earlier reports on nodal-line semimetals \cite{Ahn.2017,Carbotte.2017}.
The constant region of JDOS reflects the linear relative dispersion of the valence (green) and conduction
(yellow) bands. This dispersion is effectively one dimensional: linear in the direction perpendicular to the nodal line, while almost constant along the nodal line as well as in the $z$-direction due to quasi-2D electronic structure. Such a 1D linear dispersion gives rise to a constant JDOS.
%The constant JDOS arises from both valence and conduction bands having a linear dispersion in the 'radial' direction,
%while being approximately constant in the other two directions. JDOS thus follows the density of states
%corresponding to a linear dispersion in 1D.
The onset region reflects the deviation from this idealization due to corrugation of the nodal line and gap opening in the bulk of the Brillouin zone.
%In comparison to the JDOS, which includes the phase space information but not the transition probabilities,
%the optical conductivity reveals that the transition probabilities add a strong frequency
%dependence [see Fig.\ \ref{figure4}(b)].
We observe numerically that in all the studied materials
the low-frequency limit of the U-shaped region correlates with the
boundary of the constant JDOS. Our analysis thus does not confirm the interpretation
of Ref.~\cite{Schilling.2017}, which ascribed the spectra between 250-2500~cm$^{-1}$
to the linear bands and estimated the deviations from the nodal-line shape
due to gap opening or shift away from the Fermi level to 30~meV.
Our numerical results suggest the deviations from the perfect nodal-line structure
to be an order of magnitude bigger, largely due to corrugation of the nodal line (shift away from the Fermi level).

To summarize, the U-shape of the optical conductivity reflects the proximity to an idealized
band structure with two linearly crossing (touching) bands along a surface in the Brillouin zone, forming an
effective nodal plane.
The low-frequency boundary correlates with the deviations from this idealized picture due to
gapping of the bands or shifting the band crossings away from the Fermi level
\bibnote{Accordingly, ZrGeS is closest to the ideal nodal line system, since its low-frequency limit of the U-shaped
optical conductivity is the lowest among the studied compounds. It is interesting to note that
ZrGeS has the lowest $c$/$a$ ratio among the studied Zr$XY$ materials.}.
%The rising part of the $U$-shaped region is due to other band
%combinations. In particular, the kink/peak at the high-frequency limit of the U-shaped region has
%its origin in the transition between almost parallel bands at the $X$ and the $R$ points of the Brillouin zone
%[see Fig.\ \ref{figure4}(a)], as was identified in Ref.~\onlinecite{Habe.2018}.
%The optical spectra as well as band structure of the other members of the family reveal a similar picture.

The only exception is ZrSiTe. In this compound another pair of bands approaches the Fermi level
in the vicinity of the $X-R$ line [see Fig.\ \ref{figure4}(d)]. Besides distorting the nodal-line structure in this part
of the reciprocal space, the Fermi level is pushed away from the nodal line in the rest of the Brillouin zone.
As a result, the optical conductivity spectrum depicted in Fig.\ \ref{figure4}(e) changes its shape substantially.

%It is important to note that electronically {\it all} studied Zr$XY$ materials should be considered as {\it layered} according to our
%theoretical results, and that the chemical pressure only causes a fine-tuning of the interlayer bonding.
\subsection{Band structure of ZrXY}
In order to understand the band structure of the Zr$XY$ compounds, we have constructed a tight-binding model on the basis of Wannier orbitals with Zr $d$, Si $s$, $p$ and S (Te) $p$ characters. In this way, the valence and low-energy conduction bands are represented exactly, wheras the higher-lying conduction bands of Si $p$ and Zr $d$ character are disentangled from the rest of the band structure. Next, we have calculated a series of band structures including an increasing number of hopping processes.

In the first (upper) row of Fig.~\ref{figure5} only the nearest-neighbor Si-Si (in-plane) hopping is taken into account. Even this simple model captures the gross features of the
Zr$XY$ band structure, i.e., flat empty Zr-$d$ and occupied S(Te)-$p$ overlaid with broad
2D Si-$sp$ bands. Whereas the Si-$p_z$ bands will be removed from the Fermi level due to the inter-layer hybridization, the crossings of Si-$p_xp_ys$ bands are the precursors of the nodal-line.

%In the second row [Figs.\ \ref{figure5}(c) and (d)], all nearest-neighbor (in-plane) S(Te)-S(Te) and Zr-Zr as well as Si-Zr and Si-S(Te) hoppings are included. We note that strong Zr-Si-$p_z$ hybridization removes the $p_z$ bands from the vicinity of the Fermi level,

In the second row [Figs.\ \ref{figure5}(c) and (d)],
the nearest neighbor Si-Si, Si-Zr, and Zr-S(Te) hoppings are included. We note that strong Zr-Si-$p_z$ hybridization removes the $p_z$ bands from the vicinity of the Fermi level,
eventually forming a flat valence band between $-1$ and $-2$ eV as well as contributing to the conduction band above 5~eV.
%The S(Te)-Si bybridization is quite weak and results in a number of small gaps due to avoided band crossing.
The Zr-Si-$p_xp_ys$ hybridization is stronger. A typical shape arising from hybridization between broad and narrow bands can be observed.
The band crossing of Si-$p_xp_ys$ bands from the upper row of Fig.~\ref{figure5} are now doubled forming Si-Zr bonding and anti-bonding counterparts below and above the Zr-$d$ manifold.
The band structure obtained with unlimited hopping, depicted in Figs.~\ref{figure5}(e) and (f), exhibits only quantitative deviations from the one in the second row.

To summarize, the gross features of the electronic structure of Zr$XY$ can be understood as an overlap of broad $sp$-bands of a single
$X$ layer with much narrower Zr-$d$ (empty) and $Y$-$p$ bands (full). The precursor of the nodal line can be found in the crossing of backfolded $X$-$sp_xp_y$ bands. Hybridization with the Zr layer removes the $X$-$p_z$ bands from the vicinity
of $E_F$, whereas the $X$ $sp_xp_y$-Zr$d$ hybrid forms the nodal line.

\section{DISCUSSION}

Angle-resolved photoemission experiments on ZrSiS combined with electronic band structure calculations found a
diamond-shaped 3D-like Fermi surface due to a line of Dirac nodes \cite{Neupane.2016,Schoop.2016}. Furthermore, it was suggested
that the electronic band structure of ZrSiS contains additional Dirac-like band crossings located several hundred meV above and below
$E_F$ at the $X$ and $R$ point of the Brillouin zone. These band crossings are protected by non-symmorphic symmetry against
gapping due to the spin-orbit coupling. A comparative study on ZrSiS, ZrSiSe, and ZrSiTe proposed that the energy positions
of these band crossings depend on the $c$/$a$ ratio \cite{Schoop.2016,Topp.2016,Hosen.2017}, which may serve as a measure
for the interlayer bonding \cite{Wang.1995}. Whereas in ZrSiS the band crossings are located at 0.5 - 0.7~eV above and below
$E_F$, for layered ZrSiTe without significant interlayer bonding they are located close to
$E_F$ \cite{Topp.2016}, i.e., with increasing $c$/$a$ ratio the band crossings shift towards $E_F$.
It was also suggested that the electronic structure is very similar for the compounds Zr$XY$ ($X$=Si, Ge; $Y$=S, Se, Te)
{\it including} ZrSiTe, with only a fine tuning due to the chemical pressure effect \cite{Hu.2016,Topp.2016}.

\begin{figure}[t]
\includegraphics[width=0.5\textwidth]{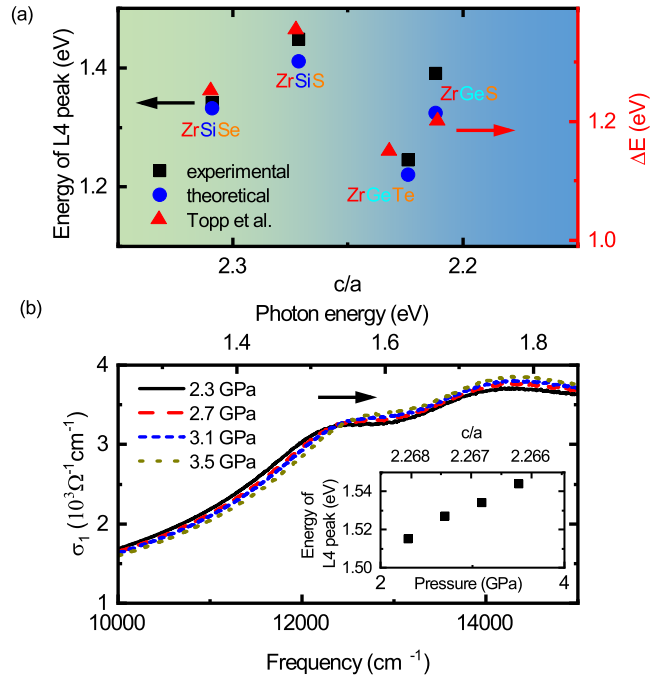}
\caption{(a) Energy position of the L4 peak as a function of $c$/$a$ ratio, obtained from the experimental and theoretical optical conductivity spectra of ZrSiS, ZrSiSe, ZrGeS, and ZrGeTe, in comparison with the energy difference $\Delta E$
between the ungapped band crossings above and below $E_F$ at the $X$ point, from Topp et al.\ \cite{Topp.2016}. Please note the offset of
0.11~eV between the left and right ordinates.
(b) Experimental optical conductivity of ZrSiS
for selected external pressures in the high-energy range. Inset: Energy position of the L4 peak in ZrSiS as a function of external
pressure. The pressure dependence of the $c$/$a$ ratio was extracted from Ref.\ \cite{Singha.2018}.}\label{figure6}
\end{figure}

According to our findings, the qualitative similarities in the electronic structure and optical conductivity only hold for compounds Zr$XY$ with a similar chemical pressure (for $c$/$a$ ratios close to 2.2 - 2.3). In stark contrast, ZrSiTe with a significantly larger
$c$/$a$ ratio [see Fig.\ \ref{figure1}(b)] and hence smaller interlayer bonding, behaves distinctly different.
In particular, the compounds ZrSiS, ZrSiSe, ZrGeS, and ZrGeTe all show the sharp L4 peak in the optical conductivity, where its energy
position depends on the specific compound and presumably on the $c$/$a$ ratio.
For a quantitative analysis, we fitted the experimental
optical conductivity spectra with the Drude-Lorentz model and obtained the energy position of the L4 peak for all studied
compounds (see suppl.). In Fig.\ \ref{figure6}(a) the energy of the L4 peak is plotted
as a function of $c$/$a$ ratio for ZrSiS, ZrSiSe, ZrGeS, and ZrGeTe, together with the positions from our theoretical results.
We find that with increasing $c$/$a$ ratio (decreasing chemical pressure) the sharp L4 peak shifts to lower energies.
This holds for ZrSiSe as compared to ZrSiS, and for
ZrGeTe as compared to ZrGeS. However, when comparing ZrSiS with ZrGeS, it seems
that a simple chemical pressure effect is not strictly given for the replacement of the carbon group element.
Apparently, the simple correlation between chemical pressure and peak position does not strictly
hold for substitutions in the $X$ square nets. This might be due to the indirect character of the chemical pressure effect induced
by the $X$ substitution, as already described in the introduction.

The L4 peak in the optical conductivity stems from transitions between almost parallel bands in the vicinity of the $X$ and the $R$ points
of the Brillouin zone, as described above. At the same $k$ points
the ungapped Dirac-like band crossings are located \cite{Schoop.2016,Topp.2016}. Therefore, the L4 peak might be related to the band crossings at the $X$ and $R$ points, respectively. In order to check this, we compare in Fig.\ \ref{figure6}(a) the energy difference $\Delta$$E$
between the ungapped band crossings above and below $E_F$ at $X$, as obtained by Topp et al.\ \cite{Topp.2016}, with the energy
of the L4 peak from the experimental and theoretical optical conductivity spectra. Apparently, there is a qualitative agreement
between the results. The position of the L4 peak in the optical conductivity spectrum may thus serve as a measure
for the energy difference between the ungapped band crossings above and below $E_F$, which are protected by non-symmorphic symmetry.

The influence of the interlayer bonding on the electronic structure is further corroborated by pressure-dependent
optical studies on ZrSiS up to 3.6~GPa. This pressure is below the critical pressure of the structural phase transition observed in Ref.\ \cite{Singha.2018}.
Generally, applying hydrostatic pressure to a material is a direct and superior way to tune the dimensionality
of a material \cite{Nagata.1998,Pashkin.2010}. For a layered material hydrostatic pressure is expected to mostly affect
the lattice parameter along the direction with the highest compressibility (at least for moderate pressures), which is the
direction perpendicular to the layers. Hence, under external pressure the distance between the layers is expected to decrease,
leading to an increase in the interlayer bonding. According to the optical conductivity of ZrSiS for selected hydrostatic pressures [see Fig.\ \ref{figure6}(b)], the L4 peak shifts to higher energy with increasing
external pressure. This finding is consistent with the observed chemical pressure
effect \bibnote{Please note that according to Ref.\ \onlinecite{Singha.2018} the effect of external pressure
on the $c$/$a$ ratio in ZrSiS is much smaller than the corresponding effect of chemical pressure. Therefore, the effect of
external pressure on the energy position of the L4 peak occurs on a much smaller energy scale as compared to the
chemical pressure effect shown in Fig.\  \ref{figure6}(a).}.

\section{CONCLUSION}

The comparative study of the optical conductivity for the compounds ZrSi$Y$ with $Y$=S, Se, Te and ZrGe$Y$ with $Y$=S, Te
revealed a similar optical conductivity profile, namely a distinct U-shape ending at a sharp peak, for all studied materials except ZrSiTe. The U-shape of the optical conductivity correlates with the nodal-line electronic structure. The low-frequency boundary of the U-region correlates with the deviations from a flat nodal line. The sharp peak at the high-frequency limit of the U-shaped region has its origin in the transitions between almost parallel bands in the vicinity of the $X$ and $R$ points of the Brillouin zone. Its energy
position may serve as a measure
for the energy difference between the ungapped band crossings above and below $E_F$ at the $X$ and $R$ points. The energy position of the peak significantly depends on the interlayer bonding of the system correlated with the $c$/$a$ ratio, which can be tuned by chemical and external pressure.
For ZrSiTe with the largest $c$/$a$ ratio the optical conductivity
profile is very different due to another pair of crossing bands in the vicinity of $E_F$, corrugating the nodal-line electronic
structure.

\begin{acknowledgments}
C.A.K. acknowledges financial support from the Deutsche Forschungsgemeinschaft (DFG), Germany, through grant no.\ KU 1432/13-1. This work was supported by the ERC Grant Agreement No. 646807 under EU Horizon 2020 (J.F.A., J.K.). The sample synthesis and characterization efforts were supported by the US Department of Energy under grant DE-SC0019068.
\end{acknowledgments}

%\bibliography{references}

%

\end{document}